\documentclass[12pt,twocolumn,preprint]{aastex6}
\usepackage{CJK}
\usepackage{color}

\usepackage{natbib}
\usepackage{hyperref}
\usepackage{cleveref}

\def\esym{$E_{\rm{sym}}(\rho)$~}
\begin{document}
\title{\bf Implications of the mass $M=2.17^{+0.11}_{-0.10}$~M$_\odot$ of PSR~J0740+6620\\ on the Equation of State of Super-Dense Neutron-Rich Nuclear Matter}
\author{Nai-Bo Zhang\altaffilmark{1} and Bao-An Li\altaffilmark{2}$^{*}$}
\altaffiltext{1}{Shandong Provincial Key Laboratory of Optical Astronomy and Solar-Terrestrial Environment,
Institute of Space Sciences, Shandong University, Weihai, 264209, China}
\altaffiltext{2}{Department of Physics and Astronomy, Texas A$\&$M University-Commerce, Commerce, TX 75429, USA\\
\noindent{$^{*}$Corresponding author: Bao-An.Li@Tamuc.edu}}

\begin{abstract}
We study implications of the very recently reported mass $M=2.17^{+0.11}_{-0.10}$~M$_\odot$ of PSR~J0740+6620 on the Equation of State (EOS) of super-dense neutron-rich nuclear matter with respect to existing constraints on the EOS based on the mass $M=2.01\pm 0.04$~M$_\odot$ of PSR~J0348+0432, the maximum tidal deformability of GW170817 and earlier results of various terrestrial nuclear laboratory experiments. The lower limit of the skewness $J_0$ measuring the stiffness of super-dense isospin-symmetric nuclear matter is raised from about -220 MeV to -150 MeV, reducing significantly its current uncertainty range. The lower bound of the high-density symmetry energy also increases appreciably leading to a rise of the minimum proton fraction in neutron stars at $\beta$-equilibrium from about 0 to 5\% around three times the saturation density of nuclear matter. The difficulties for some of the most widely used and previously well tested model EOSs to predict simultaneously both a maximum mass higher than 2.17 M$_\odot$ and a pressure consistent with that extracted from GW170817 present some interesting new  challenges for nuclear theories.
\end{abstract}
\keywords{Dense matter, equation of state, stars: neutron}
\maketitle
\newpage
\noindent{\bf Introduction:} To constrain the Equation of State (EOS) of super-dense neutron-rich nuclear matter has been a longstanding and shared goal of both astrophysics and nuclear physics \citep{Danielewicz02,LCK08,Lattimer16,Watts16,Oertel17,Ozel16,Li17,Blaschke2018}. The masses, radii and tidal deformabilities of neutron stars (NSs) are among the most promising astrophysics probes of dense neutron-rich matter EOS. In particular, the maximum mass of NSs is the best and most direct constraint on the EOS of isospin-symmetric nuclear matter (SNM) with equal numbers of neutrons and protons, while the radii and/or tidal deformability probe most directly the isospin dependent (symmetry energy) part of the EOS of neutron-rich matter in NSs \citep{LiSteiner,Zhang18}. The continuous progress in discovering and/or confirming the most massive NSs has been providing the most stringent tests of EOS models and thus improving our knowledge about the nature of and fundamental interactions in super-dense neutron-rich nuclear matter. During the last decade, the observed masses around 2M$_\odot$ of the two most massive pulsars J1614-2230 \citep{Demorest10,Arzoumanian2018} and J0348+0432 \citep{Antoniadis13} have been stimulating significantly various studies towards better understanding the EOS of super-dense neutron-rich nuclear matter.

It was just reported that by combining relativistic Shapiro delay data taken over 12.5-years at the North American Nanohertz Observatory for Gravitational Waves with recent orbital-phase-specific observations using the Green Bank Telescope, the mass of the millisecond pulsar J0740+6620 was measured to be $2.17^{+0.11}_{-0.10}$~M$_\odot$ (68.3\% credibility interval) \citep{M217}. While the error bars of this mass value are still quite large and several instances of revising down the earlier reported Shapiro delay mass measurements are causes for cautions,  as the most massive pulsar observed so far if its reported mass stays approximately unchanged, the mass of this pulsar is expected to provide more stringent constraints on the EOS of super-dense neutron-rich nuclear matter. Excited by this new discovery, using the combined constraints on the EOS based on the mass $M=2.01\pm 0.04$~M$_\odot$ of PSR~J0348+0432, the maximum tidal deformability of GW170817 and existing results of various terrestrial nuclear laboratory experiments as references, we study the following three questions:
\begin{enumerate}
\item How much better can the J0740+6620 constrain the skewness parameter used to measure the stiffness of SNM at supra-saturation densities?
\item How much better can the J0740+6620 constrain the high-density symmetry energy and the corresponding proton fraction in NSs at $\beta$-equilibrium?
\item How much tighter can the J0740+6620 constrain the pressure inside NSs predicted by the state-of-the-art nuclear many-body theories?
\end{enumerate}
We found that this new most massive NS brings about not only quantitative improvements of our knowledge about the EOS of super-dense neutron-rich nuclear matter but also qualitatively new challenges for nuclear theories. \\

\noindent{\bf The theoretical framework for inferring high-density EOS parameters from observed properties of neutron stars:} Our study is carried out within a minimum model of NSs consisting of neutrons, protons, electrons and muons (charge neutral $npe\mu$ matter) at $\beta$-equilibrium. For comparisons with the latest constraints on the EOS of super-dense neutron-rich nuclear matter available before the announcement of the mass $2.17^{+0.11}_{-0.10}$~M$_\odot$ of J0740+6620 \citep{M217}, we reuse some of our previous results obtained within the same approach and published very recently in refs. \citep{Zhang18,Zhang19}. For completeness and easy of discussions, we first summarize here the key aspects of the NS model we used. We adopted the BPS EOS  \citep{Baym71} for the outer crust and the NV EOS \citep{Negele73} for the inner crust with a crust-core transition density/pressure self-consistently determined by examining when the incompressibility of NS matter in the core becomes imaginary (the so-called thermodynamical approach), indicating the onset of cluster formations  \citep{Kubis04,Kubis07,Lattimer07,Xu09}. Unless absolutely necessary, we skip here most of the formalisms for calculating the pressure as a function of energy density used as input to solve the Tolman-Oppenheimer-Volkov (TOV) equation within the minimal model for NSs  as they are well known and can be easily found in the literature. For the purpose of inferring the high-density EOS parameters of neutron-rich nucleonic matter from astrophysical observations, the most fundamental input is the nucleon specific energy $E_b(\rho,\delta)$ in nucleonic matter with isospin asymmetry $\delta=(\rho_{\rm{n}}-\rho_{\rm{p}})/\rho$ at density $\rho$. Once the $E_b(\rho,\delta)$ is known, the pressure of NS matter at $\beta$-equilibrium
\begin{equation}\label{pressure}
  P(\rho, \delta)=\rho^2\frac{d\epsilon(\rho,\delta)/\rho}{d\rho}
\end{equation}
can be calculated numerically from the energy density $\epsilon(\rho, \delta)=\rho [E_b(\rho,\delta)+M_N]+\epsilon_l(\rho, \delta)$ where $M_N$ is the average nucleon mass and $\epsilon_l(\rho, \delta)$
is the lepton energy density. Supported by extensive studies in nuclear theory, see, e.g., ref. \citep{Bom91}, the $E_b(\rho,\delta)$ can be well parameterized as
\begin{equation}\label{PAEb}
  E_b(\rho,\delta)=E_0(\rho)+E_{\rm{sym}}(\rho)\delta^2
\end{equation}
where $E_0(\rho)$ is the nucleon specific energy in SNM and $E_{\rm{sym}}(\rho)$ is the nuclear symmetry energy.  Both the $E_0(\rho)$ and $E_{\rm{sym}}(\rho)$ can be further parameterized as
\begin{eqnarray}\label{E0para}
  E_{0}(\rho)&=&E_0(\rho_0)+\frac{K_0}{2}(\frac{\rho-\rho_0}{3\rho_0})^2+\frac{J_0}{6}(\frac{\rho-\rho_0}{3\rho_0})^3,\\
  E_{\rm{sym}}(\rho)&=&E_{\rm{sym}}(\rho_0)+L(\frac{\rho-\rho_0}{3\rho_0})+\frac{K_{\rm{sym}}}{2}(\frac{\rho-\rho_0}{3\rho_0})^2\nonumber\\
  &+&\frac{J_{\rm{sym}}}{6}(\frac{\rho-\rho_0}{3\rho_0})^3\label{Esympara}
\end{eqnarray}
where $\rho_0$ is the saturation density of SNM. As discussed in great details in refs. \citep{Zhang18}, the above expressions have the dual meanings of being Taylor expansions at small values of $\delta$ and/or $(\rho-\rho_0)/\rho_0$ on one hand, and on the other hand being purely parameterizations when they are used in very neutron-rich matter with $\delta\rightarrow 1$ at either very low or high densities significantly away from $\rho_0$. As parameterizations, mathematically they can be used at any $\rho$ and $\delta$ without the convergence issue associated with the Taylor expansions. Interestingly, the above parameterizations naturally become the Taylor expansions in the limit of $\delta\rightarrow 0$ and/or $\rho\rightarrow \rho_0$, facilitating the use of asymptotic boundary conditions for the EOS near $\rho_0$ and $\delta=0$ provided by terrestrial nuclear laboratory experiments and/or nuclear theories.

We use the Eqs. (\ref{PAEb}-\ref{Esympara}) as parameterizations and fix the $E_0(\rho_0), E_{\rm{sym}}(\rho_0), L$ and $K_0$ at their currently known most probable values from terrestrial nuclear laboratory experiments and/or nuclear theories. We then explore properties of NSs in the three-dimensional (3D) EOS parameter space of $J_0-K_{\rm{sym}}-J_{\rm{sym}}$ covering the entire space of high-density neutron-rich nucleonic matter. While in principle these parameters are completely free, the asymptotic boundary conditions of the EOS near $\rho_0$ and $\delta=0$ provide some prior knowledge about the ranges of these parameters. It is thus necessary to briefly discuss the characterizations of the EOS near the saturation point of SNM and the available constraints on them. The relevant Taylor expansion coefficients near the saturation point are determined by the $E_b(\rho,\delta)$ via $E_{\rm{sym}}(\rho) = \frac{1}{2}[\partial^2 E_b(\rho,\delta)/\partial\delta^2]_{\delta=0}$ for the symmetry energy, $L=3\rho_0[\partial E_{\rm{sym}}(\rho)/\partial\rho]|_{\rho=\rho_0}$ for the slope parameter of $E_{\rm{sym}}(\rho)$, $K_0=9\rho_0^2[\partial^2 E_0(\rho)/\partial\rho^2]|_{\rho=\rho_0}$ and $K_{\rm{sym}}=9\rho_0^2[\partial^2 E_{\rm{sym}}(\rho)/\partial\rho^2]|_{\rho=\rho_0}$ for the incompressibility of SNM and the curvature of $E_{\rm{sym}}(\rho)$, as well as $J_0=27\rho_0^3[\partial^3 E_0(\rho)/\partial\rho^3]|_{\rho=\rho_0}$ and $J_{\rm{sym}}=27\rho_0^3[\partial^3 E_{\rm{sym}}(\rho)/\partial\rho^3]|_{\rho=\rho_0}$ for the skewness of $E_0(\rho)$ and $E_{\rm{sym}}(\rho)$, respectively.
Extensive studies in both astrophysics and nuclear physics have constrained some of these coefficients to reasonably small ranges. For example, the widely accepted empirical value of $E_0(\rho_0)$ is $-15.9 \pm 0.4$ MeV \citep{Brown14}. While there are still some different opinions it is now relatively well accepted that $K_0\approx 240 \pm 20$ MeV \citep{Shlomo06,Piekarewicz10}, $E_{\rm sym}(\rho_0)=31.7\pm 3.2$ MeV and $L\approx 58.7\pm 28.1 $ MeV \citep{Li13,Oertel17,Li17}. However, the three coefficients characterizing the high-density behavior of neutron-rich matter are only very roughly known to be around $-400 \leq K_{\rm{sym}} \leq 100$ MeV, $-200 \leq J_{\rm{sym}}\leq 800$ MeV, and $-800 \leq J_{0}\leq 400$ MeV mostly based on analyses of terrestrial nuclear experiments and energy density functionals \citep{Cai17,Tews17,Zhang17}, respectively.
Given the current knowledge about the parameters discussed above, the $J_0$ is an effective parameter measuring the stiffness of SNM EOS at supra-saturation densities, which is presently the most uncertain parameter of the SNM EOS. While the $K_{\rm{sym}}$ and $J_{\rm{sym}}$ together characterize the high-density nuclear symmetry energy, they are both poorly known. Thus, constraining the $J_0-K_{ \rm{sym}}-J_{\rm{sym}}$ high-density EOS parameter space by combining observables from astrophysical observations of NSs and terrestrial experiments will help achieve the ultimate goal of determining the nature and EOS of super-dense neutron-rich nuclear matter. Compared to the widely used pieceweise polytropes for parameterizing directly the pressure as a function of energy/baryon density for super-dense NS matter, the parameterizations adopted here have explicit isospin-dependence required for inferring the high-density symmetry energy parameters and the compositions of NSs.

The most widely used technique in recent years to solve the inverse-structure problem of NSs is the Bayesian statistical inference of the posterior probability density distribution functions (PDFs) of EOS parameters, see, e.g. refs. \citep{Steiner10,Raithel17,Riley18,Lim2019,Miller19}. Our own Bayesian inferences of the EOS parameters from combined data of nuclear experiments and astrophysical observations are underway and will be reported elsewhere \citep{Xie19}. In refs. \citep{Zhang18,Zhang19}, using the parameterizations discussed above we have developed a numerical technique of inverting the TOV equation in the 3D high-density EOS parameter space for a specified NS observable, such as the radius, tidal deformability or maximum mass, or physical requirement, such as the causality condition, etc. Namely, for a given value of any observable (e.g., the upper/lower limit of certain confidence level or the most probable value), we can infer the constant surface of the observable in the 3D high-density EOS parameter space. The constant surface shows all combinations of EOS parameters necessary for giving the specified value of the observable. For the purposes of the present comparative study, inversions of the reported most probable or the 90\% upper confidence NS observables in the 3D high-density EOS parameter space are sufficient. While the boundaries of their reported confidence intervals can also be similarly inverted.\\

To this end, it is useful to discuss briefly why the observation of NS maximum mass constrain mostly the SNM EOS while the NS radii constrain mostly the density dependence of nuclear symmetry energy. For this purpose, it is sufficient to just examine the pressure of $npe$ matter before muons appear in NSs
\begin{equation}\label{pre-npe}
  P(\rho, \delta)=\rho^2[\frac{dE_0(\rho)}{d\rho}+\frac{dE_{\rm{sym}}(\rho)}{d\rho}\delta^2]+\frac{1}{2}\delta (1-\delta)\rho E_{\rm{sym}}(\rho).
\end{equation}
The density profile of isospin asymmetry $\delta(\rho)$ at $\beta$ equilibrium is uniquely determined by the
density dependence of nuclear symmetry energy \esym through the chemical equilibrium condition. Unless the \esym becomes negative leading to the isospin separation instability where SNM is unstable against being separated into regions of pure neutron matter and pure proton matter \citep{Li17}, a stiffer symmetry energy (due to a higher value of any  parameter in its parameterization)  leads to less neutron-rich matter due to the $E_{\rm{sym}}(\rho)\delta^2$ term in the nucleon specific energy in asymmetric nuclear matter. 

The first term in Eq. (\ref{pre-npe}) is the pressure $P_0$ of SNM while the last two terms are the isospin-asymmetric pressure $P_{\rm asy}$ from nucleons and electrons, separately. At the saturation density $\rho_0$, the $P_0$ vanishes and the electron contribution is also negligible leaving the total pressure determined completely by the slope of the symmetry energy. Both the $P_0$ and $P_{\rm asy}$ increase with density with rates determined by the respective density dependences of SNM EOS and symmetry energy. In the region around $\rho_0\sim 2.5\rho_0$, the $P_{\rm asy}$ dominates over the $P_0$ using most EOSs as demonstrated in Fig. 146 of ref. \citep{LCK08}. At higher densities, the SNM pressure $P_0$ dominates while the
$P_{\rm asy}$ also pays an important role depending on the high-density behaviors of nuclear symmetry energy.
The exact transition of dominance from $P_{\rm asy}$ to $P_0$ depends on the stiffnesses of both the SNM EOS and the symmetry energy. It is well known that the radii of NSs are essentially determined by the pressure at densities around $\rho_0\sim 2.5\rho_0$ \citep{Lattimer00} while the maximum masses are determined by the pressure near the central density of NSs. It is worth emphasizing that the contribution of $P_{\rm asy}$ to the total pressure at high densities can be very significant depending on both the magnitude and the slope of the symmetry energy at those densities. As indicated by Eq. (\ref{pre-npe}) and the discussions above, the symmetry energy not only affects directly the pressure but also the composition profile $\delta(\rho)$ of NS matter. The high-density symmetry energy can thus also affect significantly the maximum mass of NSs, albeit not as important as the $J_0$ parameter of the SNM EOS,  as demonstrated in Fig. 5 of ref. \citep{Zhang19}.  The above discussions explain the earlier finding \citep{LiSteiner} that fixing the incompressibility of SNM but varying the slope $L$ of nuclear symmetry energy at saturation density only changes the radii without changing the maximum mass of NSs, while fixing the symmetry energy but varying the incompressibility of SNM at saturation density only changes the maximum mass with little effect on the radii. More recently, using the same parametric EOSs as in this work, by fixing parameters characterizing the EOS of SNM and symmetry energy near the saturation density (namely, fixing the $E_0(\rho_0), E_{\rm{sym}}(\rho_0), L$ and $K_0$ at their currently known most probable values), the respective roles of the high-density SNM EOS and symmetry energy in determining the maximum mass and radii of NSs were clearly demonstrated in Fig. 5 of ref. \citep{Zhang18}. It was shown that the radii are strongly affected by the $K_{\rm{sym}}$ and $J_{\rm{sym}}$ characterizing the high-density \esym with almost no influence from the $J_0$ parameter characterizing the stiffness of SNM EOS. On the other hand, the maximum mass is mostly determined by the $J_0$ parameter. The high-density \esym starts playing a significant role on the maximum mass only when the \esym becomes super-soft leading to almost pure neutron matter in NSs. In this case, while the $\delta$ is close to 1, the  $dE_{\rm{sym}}(\rho_0)/d\rho$ is very low positively or large negatively, requiring a
much higher value of $J_0$ to support NSs of the same masses.
\\

\begin{figure}[ht]
\begin{center}
\resizebox{0.8\textwidth}{!}{
\hspace{-2cm}
  \includegraphics{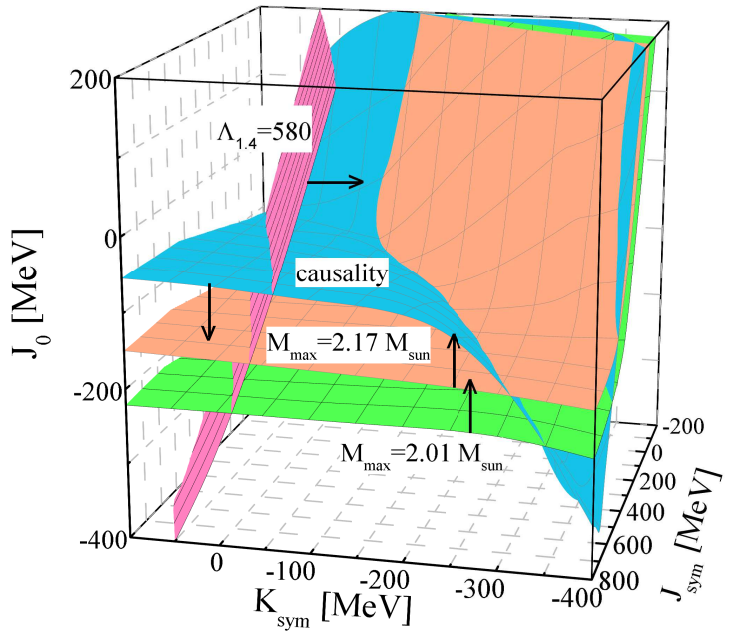}
  }
  \vspace{-2cm}
  \caption{(color online) Constant surfaces of NS maximum mass of $M_{\rm max}=2.01$ M$_\odot$ and $M_{\rm max}=2.17$ M$_\odot$ as well as the maximum tidal deformability
  $\Lambda_{1.4}=580$ (90\% confidence level) for canonical NSs and the causality condition, respectively, in the $J_0-K_{\rm sym}-J_{\rm sym}$ parameter space for high-density neutron-rich nuclear matter.
  }\label{mass217}
\end{center}
\end{figure}
\noindent{\bf  J0740+6620 implications on the stiffness of super-dense isospin-symmetric nuclear matter:} Shown in Fig. \ref{mass217} are the constant surface of the NS maximum mass $M_{\rm max}=2.01$ M$_\odot$ in comparison with that of $M_{\rm max}=2.17$ M$_\odot$ in the $J_0-K_{\rm sym}-J_{\rm sym}$ parameter space for high-density neutron-rich nuclear matter.
Every point in this 3D space represents a unique EOS defined by Eqs.  (\ref{pressure}-\ref{Esympara}).
The causality surface is defined as where the speed of sound becomes equal to the speed of light at the central density of the most massive NS allowed by a given EOS \citep{Zhang19}. It restricts the maximum value of the skewness parameter $J_0$. As it was discussed in detail in ref. \citep{Zhang19}, corresponding to the $J_0$ value at each point on the causality surface there is a NS maximum mass. The maximum masses on the causality surface reach an EOS-independent limit, i.e., the absolutely maximum mass of NSs at about 2.40 M$_\odot$ \citep{Zhang19}. More recently, using their latest and improved version of the zero temperature quark-hadron crossover EOS, QHC19, Baym et al. \citep{Baym19} found an absolutely maximum mass of 2.35 M$_\odot$ at the casual limit consistent with that found in ref. \citep{Zhang19}. Compared to these predicted absolutely maximum masses of NSs, the finding of a NS with 2.17 M$_\odot$ is not surprising and there is still a large room to go.
In the context of this work, it is also interesting to note here that many interesting studies have been carried out to estimate the maximum mass of neutron stars using signals from the GW170817-GRB170817A-AT2017gfo events. For example, it has been reported that the NS maximum mass is $2.17$ M$_\odot$ at 90\% confidence~\citep{Mar17}, $2.16^{+0.17}_{-0.15}$ M$_\odot$ at 90\% confidence~\citep{Rez18}, $2.16-2.28$ M$_\odot$ when the ratio of the maximum mass of a uniformly rotating max
neutron star (the supramassive limit) over the maximum mass of a nonrotating star is within $1.2\leq\beta \leq1.27$~\citep{Ruiz18}, $2.15-2.25$ M$_\odot$ after reducing effects of gravitational-wave emission, long-term neutrino emission, ejected mass, and rotation from the total mass of GW170817 $2.73\sim2.78$ M$_\odot$~\citep{Shibata2017}, $2.18$ M$_\odot$ (2.32 M$_\odot$ when pairing is considered) using the upper limit of $\Lambda_{1.4}=800$ (90\% confidence)~\citep{Zhou2018} and $2.3$ M$_\odot$ considering the conservation laws of energy and angular momentum self-consistently~\citep{Shibata2019}.
While the fate of GW170817 is not completely known observationally, all these studies point to a maximum mass less than $2.4$ M$_{\odot}$ consistent with the absolutely maximum mass predicted in refs. \citep{Zhang19,Baym19}.

Of course, the causality surface depends on the high-density symmetry energy parameters $K_{\rm sym}$ and $J_{\rm sym}$. For example, when these two parameters are large and positive, the symmetry energy is stiff/high, the high-density NS matter at $\beta$-equilibrium is energetically more favorable to have smaller $\delta$ values due to the $E_{\rm sym}(\rho)\cdot \delta^2$ term in the nucleon specific energy. Effects of the symmetry energy on the pressure and its slope (speed of sound squared) are relatively small compared to the contributions from SNM EOS.  Consequently, the constant surfaces of the maximum mass and causality surface are rather flat. However, when the $K_{\rm sym}$ and $J_{\rm sym}$ are small or become negative, the high-density symmetry energy is rather soft/low, and the high-density NS matter is energetically more neutron-rich with larger values of $\delta$, which makes the effects of symmetry energy more obvious. Thus, the contribution of soft/low symmetry energy to the pressure in eq. (1) is small although the value of $\delta$ may be high, to support the same maximum mass or reach the same causality surface the SNM EOS is required to be more stiff with larger $J_0$ values. As a result, the constant surfaces of the maximum mass and causality both turn up to higher $J_0$ values when both the $K_{\rm sym}$ and $J_{\rm sym}$ become negative, leading to super-soft symmetry energies at the right-back corner. More quantitatively, when the $K_{\rm sym}$ and $J_{\rm sym}$ are varied in the allowed full range, the $J_0$ parameter on the
surface of a constant maximum mass $M_{\rm max}=2.01$ M$_\odot$ varies from about $-220$ MeV to $+200$ MeV. Thus, besides the stiffness of SNM EOS, the high-density symmetry energy also plays a significant role in determining the maximum mass of NSs. Compared to the prior range of $-800 \leq J_{0}\leq 400$ MeV from terrestrial nuclear experiments and predictions of nuclear many-body theories, it is interesting to note that properties of NSs reduce this range significantly. In particular, the maximum mass of NSs sets a rather stringent lower limit for the value of $J_0$.
It is seen that the two surfaces with maximum masses of $M_{\rm max}=2.01$ M$_\odot$ and $M_{\rm max}=2.17$ M$_\odot$ are approximately parallel in the whole space. In the front region where the symmetry energy is stiff/high, the lower limit of the skewness $J_0$ increases by approximately 47\% from about $-220$ MeV to $-150$ MeV when the maximum mass raises by about 8\% from 2.01 M$_\odot$ to 2.17 M$_\odot$. Thus, the  mass of J0740+6620 pushes up the lower limit of the skewness parameter $J_0$ of SNM significantly.\\

\begin{figure}[ht]
\begin{center}
\resizebox{0.56\textwidth}{!}{
\hspace{-0.5cm}
  \includegraphics[width=5.cm]{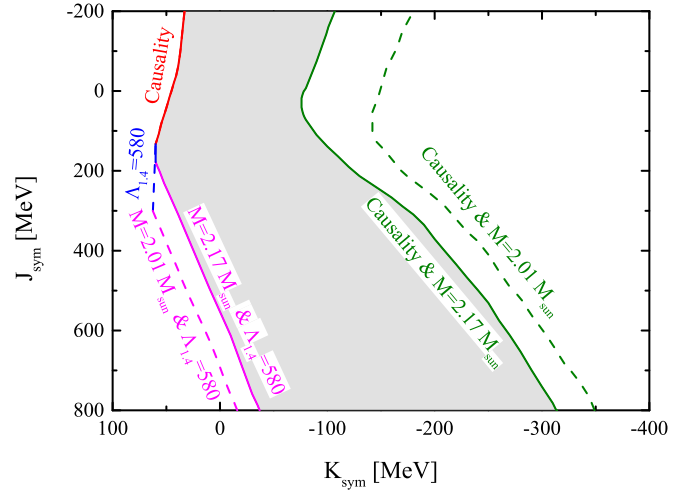}
  }
   \caption{(color online) The boundaries of the high-density symmetry energy parameter plane determined by the crosslines of the constant surfaces shown in Fig. \ref{mass217}. The shadowed range corresponds to the parameters allowed.}\label{Esym-b}
\end{center}
\end{figure}
\begin{figure*}[ht]
\begin{center}
  \resizebox{0.48\textwidth}{!}{
   \includegraphics[width=17.cm,height=14cm]{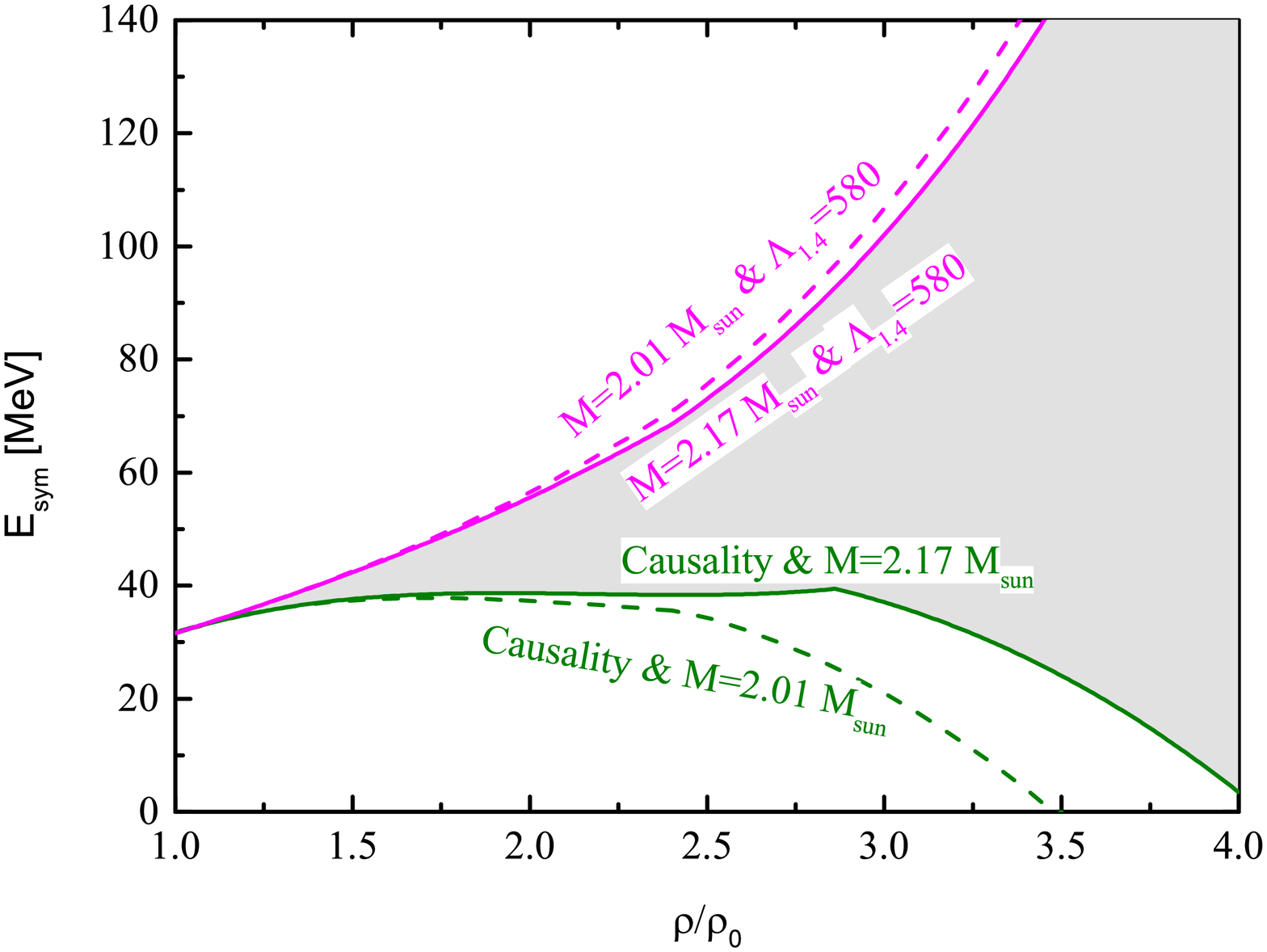}
   }
   \resizebox{0.48\textwidth}{!}{
 \includegraphics[width=17.cm,height=14cm]{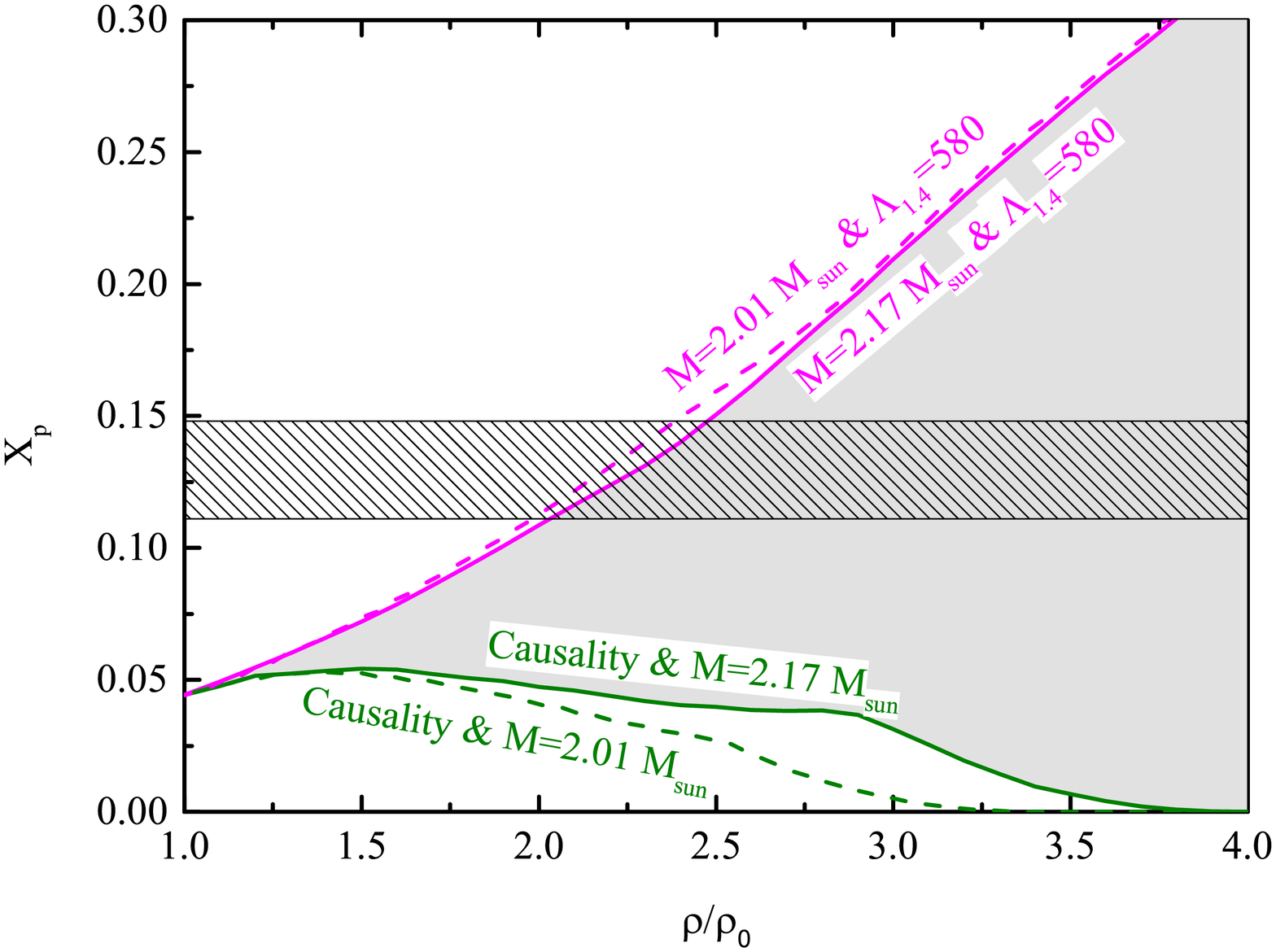}
}
  \caption{(color online) Nuclear symmetry energy in the supra-saturation density region (left) and the corresponding proton fraction in NSs at $\beta$-equilibrium (right). The horizontal band
  between 11.1\% to 14.8\% is the direct URCA limit for fast cooling of proto-NSs.
   }\label{xfraction}
\end{center}
\end{figure*}

\noindent{\bf J0740+6620 implications on the high-density symmetry energy and proton fraction in NSs:} The dimensionless tidal deformability $\Lambda$ of NSs is related to
the compactness parameter $\beta\equiv R/M$ and the Love number $k_2$ through
$
\Lambda = \frac{2}{3}\frac{k_2}{\beta^5}.
$
The $k_2$ is obtained from solving differential equations coupled to the TOV equation \citep{Hinderer08,Hinderer10,Fattoyev13}. The LIGO and VIRGO Collaborations \citep{LIGO18} found the maximum tidal deformability of canonical NSs is about $\Lambda_{1.4}=580$ (90\% confidence level). It is seen from Fig. \ref{mass217} that the constant surface of $\Lambda_{1.4}=580$ is almost vertical. This feature indicates that the skewness of SNM EOS has little effect on the tidal deformability. On the other hand, the latter has been found to depend sensitively on the high-density symmetry energy \citep{Malik2018,Plamen3,Zhang19b,Carson2019}. It is seen that the constant surface $\Lambda_{1.4}=580$ sets an observational boundary for the 3D high-density EOS parameter space from the left. While the causality surface crosses with the $\Lambda_{1.4}=580$ surface on the upper west and with the surface of the maximum mass on the lower east. Obviously, depending on whether the NS maximum mass is 2.01 or 2.17 M$_{\odot}$, the locations of the boundaries in the $K_{\rm sym}$-$J_{\rm sym}$ plane are different. Shown in Fig. \ref{Esym-b} are the boundaries of $K_{\rm sym}$-$J_{\rm sym}$ plane determined by the crosslines of the constant surfaces shown in Fig. \ref{mass217}. The increased NS maximum mass from 2.01 to 2.17 M$_{\odot}$ shrinks the allowed $K_{\rm sym}$-$J_{\rm sym}$ plane appreciably. It is probably necessary to mention that the reported lower limits of the
tidal deformability from analyzing GW170817 by different groups remain controversial. The lower limit of about 70 reported by the LIGO+VIRGO Collaborations \citep{LIGO18} falls outside the crossline between the causality surface and the  2.01 M$_{\odot}$ surface. It thus does not provide any useful constraint on the EOS as we discussed in detail in ref. \citep{Zhang19}.

Shown in Fig. \ref{xfraction} are the upper and lower boundaries of nuclear symmetry energy at supra-saturation densities using the constraints on the boundaries of $K_{\rm sym}$-$J_{\rm sym}$ in Fig.  \ref{Esym-b}. Clearly, the mass of J0740+6620 raises the lower boundary of the high-density symmetry energy and the corresponding proton fraction at $\beta$-equilibrium appreciably.
More quantitatively, the minimum proton fraction increases from about 0 to 5\% around three times the saturation density of nuclear matter when the maximum mass of NSs is increased from 2.01 to 2.17 M$_{\odot}$.  We also notice that the upper boundary from the crosslines of the maximum mass and tidal deformability is only slightly changed. As a reference, in the $npe\mu$ matter the threshold proton fraction $x^{DU}_p$ necessary for the fast cooling through the so-called direct URCA process (DU) to occur is
$
x^{DU}_p=1/[1+(1+x_e^{1/3})^3].
$ Its value is between 11.1\% to 14.8\% for the electron fraction $x_e\equiv \rho_e/\rho_p$ between 1 and 0.5 \citep{Klahn}.  This range is indicated by the horizontal band in the right window of Fig. \ref{xfraction}. It is seen that the reported variation of the NS maximum mass is not large enough to improve our knowledge about when/where/whether the direct URCA process can happen or not. \\

\begin{figure}[ht]
\begin{center}
  \resizebox{0.49\textwidth}{!}{
  \includegraphics[width=12cm]{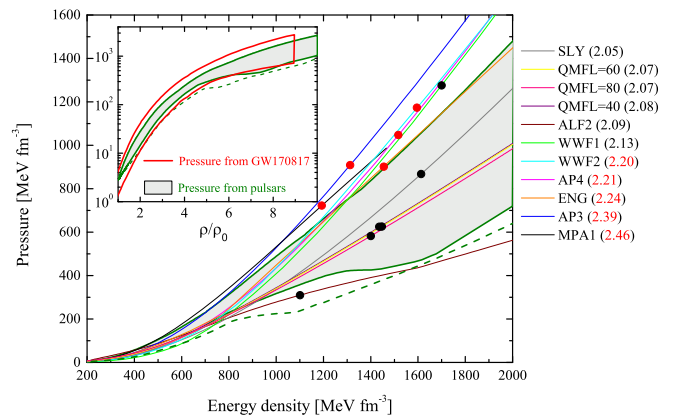}
  }
    \caption{(color online) The pressure as a function of baryon/energy density in NS matter at $\beta$-equilibrium we extracted from properties of NSs (Pressure from pulsars) shown in Fig. \ref{mass217} in comparison with the one extracted from GW170817 by the LIGO and VIRGO Collaborations at 90\% confidence level (red boundaries in the inset) and predictions of 11 EOS models. The maximum mass predicted by each model
 is indicated in the parenthesis following the model name. The black or red symbols indicate the maximum pressure and energy density reached in NSs having the maximum mass predicted in the specified model. The dashed lines correspond to the constrained lower limit of pressure obtained using $M_{\rm max}=$ 2.01 M$_{\odot}$.}\label{p-e}
\end{center}
\end{figure}

\noindent{\bf J0740+6620 implications on the pressure in NSs and nuclear theories:} Next we study how the restricted EOS parameter space constrains the pressure in NSs at $\beta$-equilibrium.
Shown in the inset of Fig. \ref{p-e} is the pressure as a function of baryon density in NS matter at $\beta$-equilibrium we extracted from pulsar properties in comparison with the one extracted from GW170817 by the LIGO and VIRGO Collaborations \citep{LIGO18} (red boundaries). The pressure as a function of energy density we extracted is compared with predictions of 11 EOS models in the main frame of Fig. \ref{p-e}. The dashed lines correspond to the constrained lower limit of pressure using $M_{\rm max}=$ 2.01 M$_{\odot}$. The 11 EOSs are representatives of predictions based on state-of-the-art nuclear many-body theories.  More specifically, they are the ALF2 of Alford et al. \citep{ALF2} for hybrid (nuclear + quark matter) stars,  AP3 and AP4 of Akmal and Pandharipande \citep{AP34}, ENG of Engvik et al. \citep{ENG}, MPA1 of Muther, Prakash and Ainsworth \citep{MPA1}, SLy of Douchin and Haensel \citep{SLy}, WWF1 and WWF2 of Wiringa, Fiks and Fabrocini \citep{WFF}, the QMFL40, QMFL60 and QMFL80 are based on the Quark Mean Field model with L=40, 60 and 80 MeV \citep{QMF}, respectively. As indicated by the values in the parenthesis following the model name of these EOS models in Fig. \ref{p-e}, while they all predicted NS maximum masses higher than 2.01 M$_{\odot}$, only five of them have maximum masses higher than 2.17 M$_{\odot}$.

Using the formalisms presented earlier, the pressure in NSs at $\beta$-equilibrium can be calculated at every point inside the restricted EOS space. The lower/upper boundaries of the pressure satisfying all constraints are obtained by finding the lowest/highest pressures at a given baryon/energy density. While the upper boundary is basically determined by the causality surface shown in Fig. \ref{mass217},  the lower boundary is mainly determined by the maximum mass of NSs with their proton fractions determined consistently by the symmetry energy of each EOS used. It is worth noting that at a given energy density, some model pressures keep increasing while some others have already increased to their maximum energy densities allowed by the constraints. Consequently, in plotting the lowest pressure as a function of baryon/energy density, some softening may appear along the lower boundary of the pressure \citep{Zhang19}. 

Several interesting observations can be made from the results shown in Fig.\ \ref{p-e}: (1) the
pressure from analyzing the GW170817 event by the LIGO and VIRGO Collaborations and what we extracted from inverting the NS properties agree very well. (2) the change of NS maximum mass from 2.01 M$_{\odot}$ to 2.17 M$_{\odot}$ moves up the lower boundary of pressure slightly. (3) Except one EOS, all other 10 EOSs predict pressures above the lower bound. Among the 10 EOSs, 4 of them having the maximum masses above 2.01 M$_{\odot}$ but less than 2.17 M$_{\odot}$ fall inside the constrained pressure band, while all 5 EOSs having the maximum masses higher than 2.17 M$_{\odot}$ run out of the pressure band at higher energy densities. We emphasize here that on the pressure-energy density plot, a higher pressure above the constraining band does not necessarily mean that the corresponding EOS is acausal (since the speed of sound is determined by the slope of the pressure with respect to energy density). In fact, all of the 11 EOSs considered here are causal in the energy density range considered. Interestingly, none of the 11 EOSs examined here falls entirely into the constrained pressure band while also supports NSs with masses higher than 2.17 M$_{\odot}$.  This obvious tension thus brings about a new challenge for nuclear theories. \\

\noindent{\bf Summary:} In summary, the exciting report of the mass $M=2.17^{+0.11}_{-0.10}$~M$_\odot$ of PSR~J0740+6620 not only helps improve quantitatively our knowledge about the EOS of super-dense neutron-rich nuclear matter but also creates some new challenges for nuclear theories.  In particular,  the lower limit of the skewness $J_0$ increases by approximately 47\% from about $-220$ MeV to $-150$ MeV, thus reducing significantly the range of this most uncertain parameter of super-dense nuclear matter when the NS maximum mass raises by about 8\% from 2.01 M$_\odot$ to 2.17 M$_\odot$. Moreover, the lower bound of the high-density symmetry energy also increases appreciably leading to a rise of the minimum proton fraction in NSs at $\beta$-equilibrium from about 0 to 5\% around three times the saturation density of nuclear matter. Furthermore, the observed new maximum mass of NSs together with the
pressure in NSs extracted from GW170817 provide so far the strongest test of the EOS models. Obviously, the difficulties for some of the most widely used and previously well tested model EOSs to predict simultaneously both a maximum mass higher than 2.17 M$_\odot$ and a pressure consistent with that extracted from analyzing the GW170817 event present some interesting new challenges for nuclear theories.

A major caveat of our results is that the observational ``data" and physics conditions we used have different confidence levels. The upper limit of tidal deformability from GW170817 is at 90\% confidence level, the two central values of the maximum masses are from measurements at 68\% confidence level while the theoretical causality surface is the absolute upper limit (i.e., at 100\% confidence level). The use of these inputs with mixed confidence levels in deriving the boundaries of the EOS parameters makes it statistically difficult to quantify the uncertainties of some of our results. Nevertheless, assuming the reported most probable mass M=2.17M$_\odot$ for PSR~J0740+6620 is equally reliable as that of M=2.01M$_\odot$ for J0348+0432, the results of our comparative studies are still scientifically sound and useful.\\

\noindent{\bf Acknowledgments:} We thank Wen-Jie Xie for very helpful discussions. This work is supported in part by the U.S. Department of Energy, Office of Science, under Award Number DE-SC0013702, the CUSTIPEN (China-U.S. Theory Institute for Physics with Exotic Nuclei) under the US Department of Energy Grant No. DE-SC0009971,
the China Postdoctoral Science Foundation (No. 2019M652358) and the Fundamental Research Funds of Shandong University.

\end{document}